\documentclass[sort&compress]{aipproc}
  
\layoutstyle{6x9}

\usepackage{subfigure}

\begin{document}

\title{Non--linear shock acceleration and high energy gamma rays from clusters of galaxies}

\author{Stefano Gabici}{
  address={Dipartimento di Astronomia e Scienza dello Spazio, Largo E. Fermi 5, 50125 Firenze, Italy \\ and Alexander von Humboldt Fellow. Present address: Max--Planck--Institut fuer Kernphysik, Saupfercheckweg 1, Heidelberg 69117, Germany},
  altaddress={INAF/Osservatorio Astrofisico di Arcetri, Largo E. Fermi 5, 50125 Firenze, Italy}
}

\author{Pasquale Blasi}{
   address={INAF/Osservatorio Astrofisico di Arcetri, Largo E. Fermi 5, 50125 Firenze, Italy}
}

\begin{abstract}
Merger and accretion shocks in clusters of galaxies can accelerate particles via first order Fermi process. Since this mechanism is believed to be intrinsically efficient, shocks are expected to be modified by the backreaction of the accelerated particles. Such a modification might induce appreciable effects on the non--thermal emission from clusters and a suppression of the heating of the gas at strong shocks. Here we consider in particular the gamma ray emission and we discuss the capability of Cherenkov telescopes such as HESS to detect clusters at TeV energies.
\end{abstract}

\maketitle


\section{Introduction}

Diffuse synchrotron radio halos are observed in a growing number of rich clusters of galaxies \cite{gigia}. This tells us that a population of relativistic electrons and a $\sim \mu G$ magnetic field are present in the intracluster medium (ICM).
Even if the existence of relativistic electrons is firmly proved by observations, the origin of these particles and the mechanisms through which they are accelerated are still unknown. 
Moreover, since the majority of the acceleration mechanisms at work in astrophysical sources accelerate protons as well, it is generally assumed that also an hadronic cosmic ray (CR) component is present in the ICM.
From X--ray observations we also know that clusters contain a considerable amount of baryons in the form of a hot diffuse gas \cite{sarazin}.
These facts led Dennison \cite{dennison} to propose that clusters might be gamma ray sources, due to the decay of neutral pions produced in the inelastic interactions between CR protons and thermal protons in the ICM.
This possibility was extensively studied after the discovery that relativistic protons remain diffusively confined within the cluster volume for cosmological times, without losing their energy \cite{bbp,vab}.
As a consequence, both the probability of having inelastic proton proton collisions and the related gamma ray emissivity are expected to be enhanced.
Other contributions to the total gamma ray emission come from 
relativistic electrons (for a review see Blasi, this conference). 

To date, clusters of galaxies have not been detected yet in gamma rays, neither at GeV \cite{olaf} nor at TeV energies. 
A number of theoretical predictions of the gamma ray luminosity of clusters have been recently proposed.
Different CR sources have been considered, including shocks in the ICM \cite{io3,dermer,uri,bbp,minia,colablasi,blasi01}, radio galaxies \cite{ensslin,bbp,colablasi,pfrommer} or starburst galaxies \cite{vab,voelk,bbp}.
The perspective for detection appears promising for future space--borne instruments such as GLAST and AGILE \cite{io3,dermer,uri,voelk,minia}.
Unfortunately, less solid conclusions can be drawn for Cherenkov telescopes, due to the extended size of the emitting region \cite{io3,voelk}.

In all the above mentioned papers, the spectra of CRs accelerated at shocks in the ICM have been calculated in the test--particle regime, namely, neglecting the CR backreaction onto the shock structure (but see also \cite{kang2}).
However, shock acceleration is believed to be an efficient mechanism and, as a consequence, the effects of CR pressure should be included in the calculations \cite{druryNL}.

Here we use the approach presented in \cite{NLpasquale} to study the relevance of these non linear effects in large scale shocks that naturally form during the process of large scale structure formation.
We show that ICM heating might be strongly suppressed at strong shocks, which are likely to convert a great fraction of the total shock kinetic energy into CRs.
Moreover, the spectra of the accelerated particles are no longer power laws, as in the test--particle approach, and this could affect our estimates of the gamma ray emission from clusters and our predictions about the detectability of these objects with new generation instruments.
In particular, for the case of ground based Cherenkov telescopes such as HESS, an accurate determination of the instrument sensitivity for extended sources is needed in order to make firm claims.

A subjective view of future perspectives in this field of research is given in the last section. We believe that an approach based on a combination of high resolution cosmological simulations and semi--analytical models for particle acceleration might be the best way to study the effects of CR acceleration on the thermal and non thermal properties of clusters.

\section{Non linear shock acceleration in the ICM}

As seen in the previous section, shocks in the ICM might be the sources of CRs in the cluster volume. 
The typical velocities of merger and accretion shocks in rich clusters are of the order of $\sim 1000 \, km/s$ and their size is $\sim 1 \, Mpc$ or more. If a $\mu G$ magnetic field is present in the shock region, protons can be accelerated up to energies of at most $\sim 5 \times 10^{19} eV$ \cite{norman,kang1} (but see \cite{ostro}, in which a smaller value $\sim 10^{17} eV$ has been found).
The determination of the shock Mach number $\cal M$ is of crucial importance in order to calculate the spectral shape of the accelerated particles. In the test--particle regime, particle spectra are simple power laws in momentum $p^2 f(p) \propto p^{-\alpha}$ with slope related to the shock Mach number through the well known relation: $\alpha = 2 ({\cal M}^2+1)/({\cal M}^2-1)$.
This means that particles accelerated at weak shocks have extremely steep spectra, while an asymptotic value $\alpha = 2$ is obtained for high Mach numbers.

Merger shocks propagating in the cluster cores are expected to have low Mach numbers, due to the very high temperature of the ICM \cite{io1,dermer}. This idea is also supported by a few X--ray observations (e.g. \cite{xrays}).
On the contrary, accretion shocks are believed to be strong since they propagate in the external cold medium and their Mach numbers could be as high as $\sim 10 - 100$, if the upstream temperature varies in the range $10^4 - 10^6 K$.

In \cite{io1} we evaluated the test--particle spectra of particles accelerated at merger shocks, assuming that a small fraction $\sim 10\%$ of the shock kinetic energy is converted into CRs.
An alternative approach is based on the so called thermal leakage recipe for injection \cite{TL}. According to this model, the post--shock gas is assumed to thermalize and only the particles in the tail of the Maxwellian distribution are allowed to recross the shock and to take part in the acceleration process. The threshold above which particles are injected is expressed as a multiple of the downstream particle thermal momentum: $p_{inj} = \xi p_{th}$, where $\xi$ is substantially a free parameter, which depends on the shock microphysics.
Although the thermal leakage approach is more satisfactory from a physical point of view, it leads to inconsistencies if particle spectra are assumed to be the nice power laws predicted by linear theory, as done in \cite{minia}.
In particular, for strong shocks (${\cal M} \sim 10$ or more) the ratio between the pressure carried by CRs and the thermal pressure comes out to be of order unity, clearly violating the test--particle assumption. For this reason, the spectral normalization has to be adjusted in a somewhat arbitrary way.
All these things seem to suggest that accurate results on particle acceleration, especially at strong accretion shocks, can be obtained only by means of non--linear calculations.

A common feature of the modified shocks is that the spectrum of accelerated particles gets progressively flatter at higher momenta, at odds with test--particle model predictions.
For strong shocks the high energy spectrum reach an asymptotic form $f(p)p^2 \propto p^{-3/2}$ and the bulk of the CR energy is carried by a few particles having the maximum momentum.
These particles can leave the system, carrying away a non negligible fraction of the total energy (e.g. \cite{simple}).
Moreover, if a great fraction of the total shock kinetic energy is converted into CRs, the heating of the gas can be appreciably suppressed \cite{druryNL}.

\begin{figure}[t]
\hfill
\begin{minipage}[t]{.5\textwidth}
\begin{center}
\includegraphics[width=.95\textwidth]{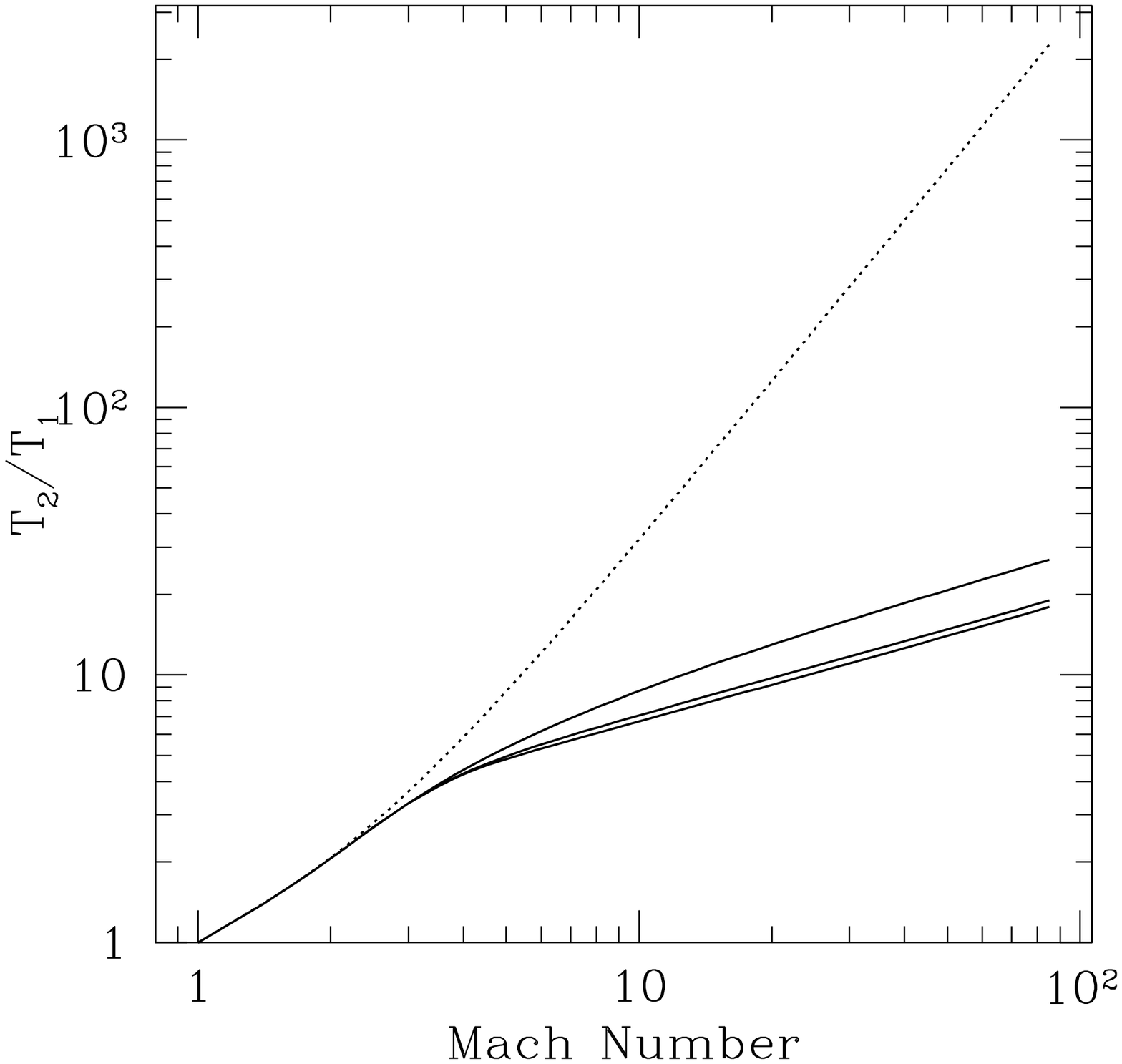}
\end{center}
\end{minipage}
\hfill
\begin{minipage}[t]{.5\textwidth}
\begin{center}
\includegraphics[width=.95\textwidth]{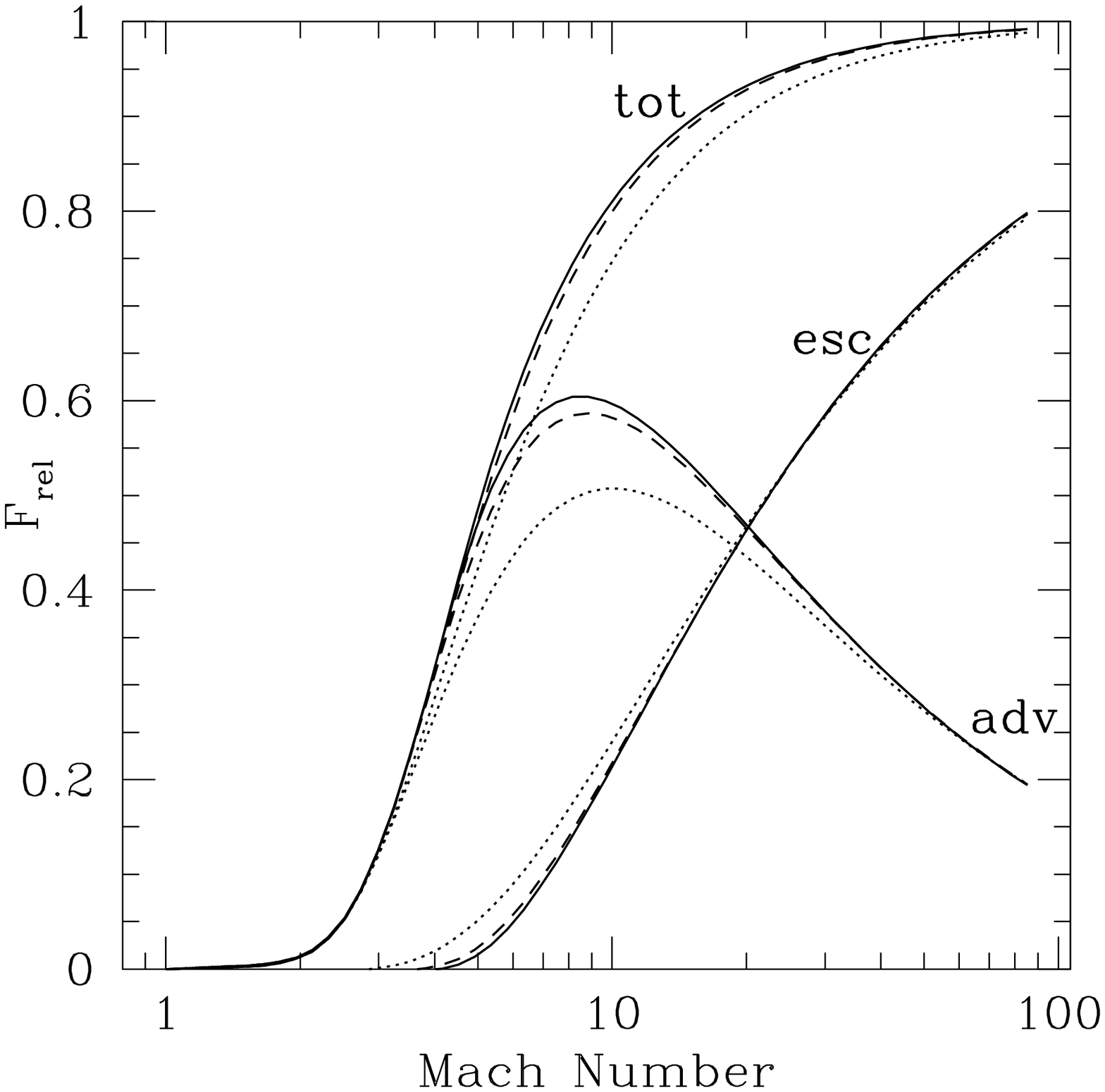}
\end{center}
\end{minipage}
\hfill
\caption{{\bf Left panel:} ratio between the downstream and upstream temperature at a shock. The dotted line represents the linear case while the solid lines are the results of non--linear calculations. Different values of the maximum momentum have been adopted ($p_{max} = 10^2, 10^6, 5\times10^{10} GeV$ from top to bottom. {\bf Right panel:} acceleration efficiency as a function of the shock strength. The total efficiency is shown, together with the two contributions from escaping particles and particles advected downstream. $p_{max}$ is equal to $10^2$ (dotted lines), $10^6$ (dashed) and $5\times10^{10} GeV$ (solid).}
\end{figure}

All these issues can be discussed with the help of Figure 1.
The dotted line in the left panel represents the ratio between the downstream and upstream temperature for an unmodified shock. The curve shows the canonical behaviour $\propto {\cal M}^2$ at high Mach numbers. The solid lines refer to a modified shock with velocity $1000 \, km/s$ and different maximum momenta of the accelerated particles (see caption). Calculations have been carried on according to \cite{NLpasquale} and adopting the thermal leakage model for particle injection with parameter $\xi \sim 3.5$ (this roughly corresponds to inject a fraction $10^{-4}$ of the particles that cross the shock)\footnote{It is known that the problem of particle acceleration at modified shocks admits multiple solutions \cite{druryNL}. However, if the thermal leakage recipe is used to describe injection, multiple solutions disappear for the shock parameters considered here. For details see Blasi, Gabici \& Vannoni, in preparation.}. It is clear that, while for very weak shocks (${\cal M} < 3-4$) there are only minor deviations from the linear predictions, for high Mach number shocks the heating of the gas can be strongly suppressed.

The acceleration efficiency is defined as the fraction of the shock kinetic energy flux that goes into accelerated particles. It is plotted in Figure 1 (right panel) for the same sets of parameters considered in the left panel.
The total acceleration efficiency, labelled with {\it tot}, is the sum of the contribution of accelerated particles that leave the system (label {\it esc}) and the ones that are advected downstream (label {\it adv}). The fraction of energy carried away by particles with maximum momentum strongly increase with the Mach number, while the CR energy advected downstream has a peak in correspondence of ${\cal M} \sim 8$. The total efficiency saturates to 1 for very strong shocks.
This means that, even if strong shocks in principle can have efficiency of order unity, the fraction of the shock kinetic energy converted into CRs that are advected downstream and stored in the cluster volume can be much smaller, depending on the exact value of the Mach number.

To conclude, it must be stressed that even a small amount of additional gas heating in the upstream fluid (e.g. Alfv\'en waves damping, acoustic instabilities) can dramatically change the situation, making the shocks less modified \cite{druryNL}.
Since we neglect these effects, our results should be considered as somewhat extreme situations. We will discuss this issue in a forthcoming paper.

\section{High energy gamma ray emission}

\begin{figure}[t]
\hfill
\begin{minipage}[t]{.5\textwidth}
\begin{center}
\includegraphics[width=.95\textwidth]{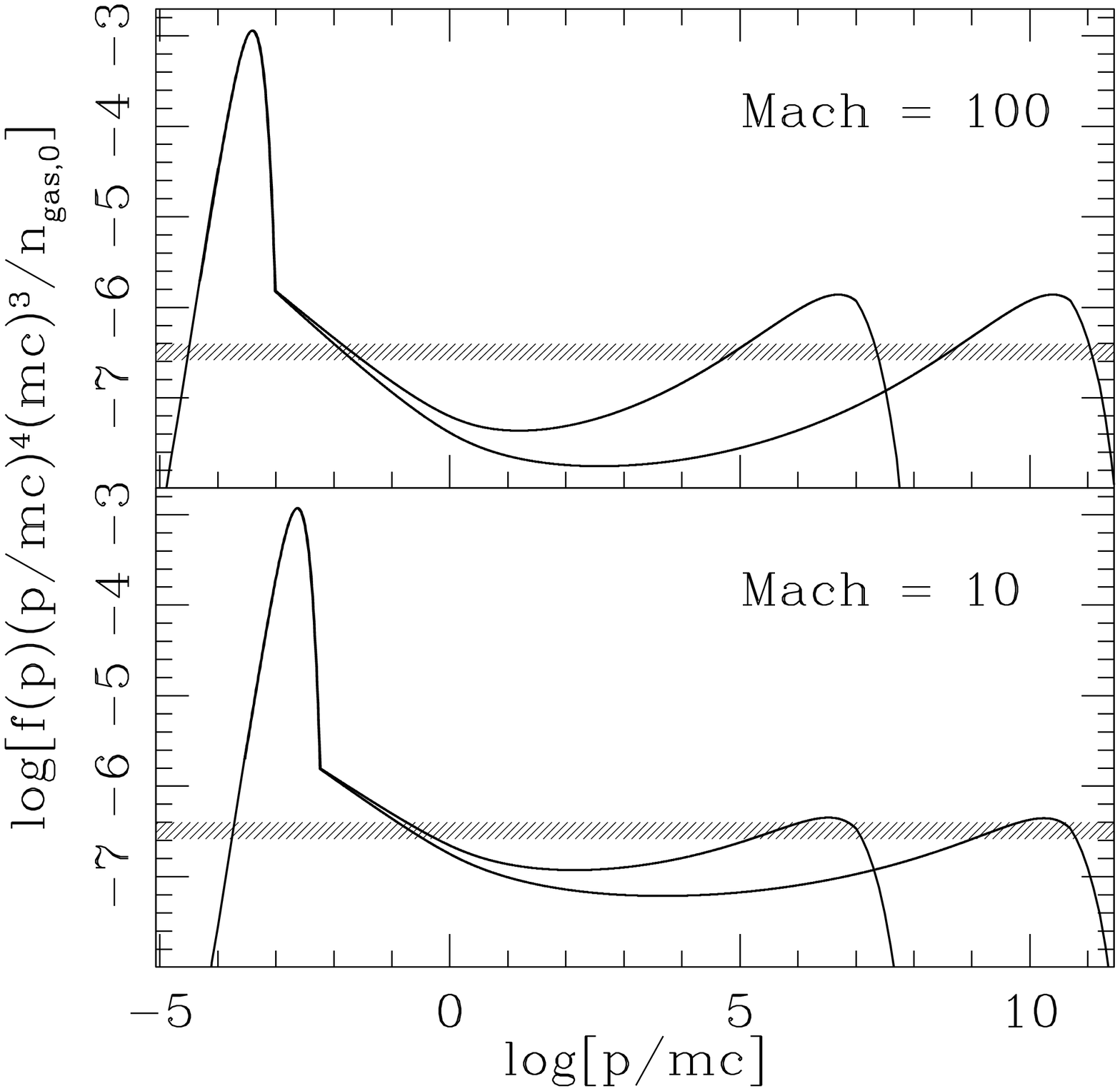}
\end{center}
\end{minipage}
\hfill
\begin{minipage}[t]{.5\textwidth}
\begin{center}
\includegraphics[width=1.\textwidth]{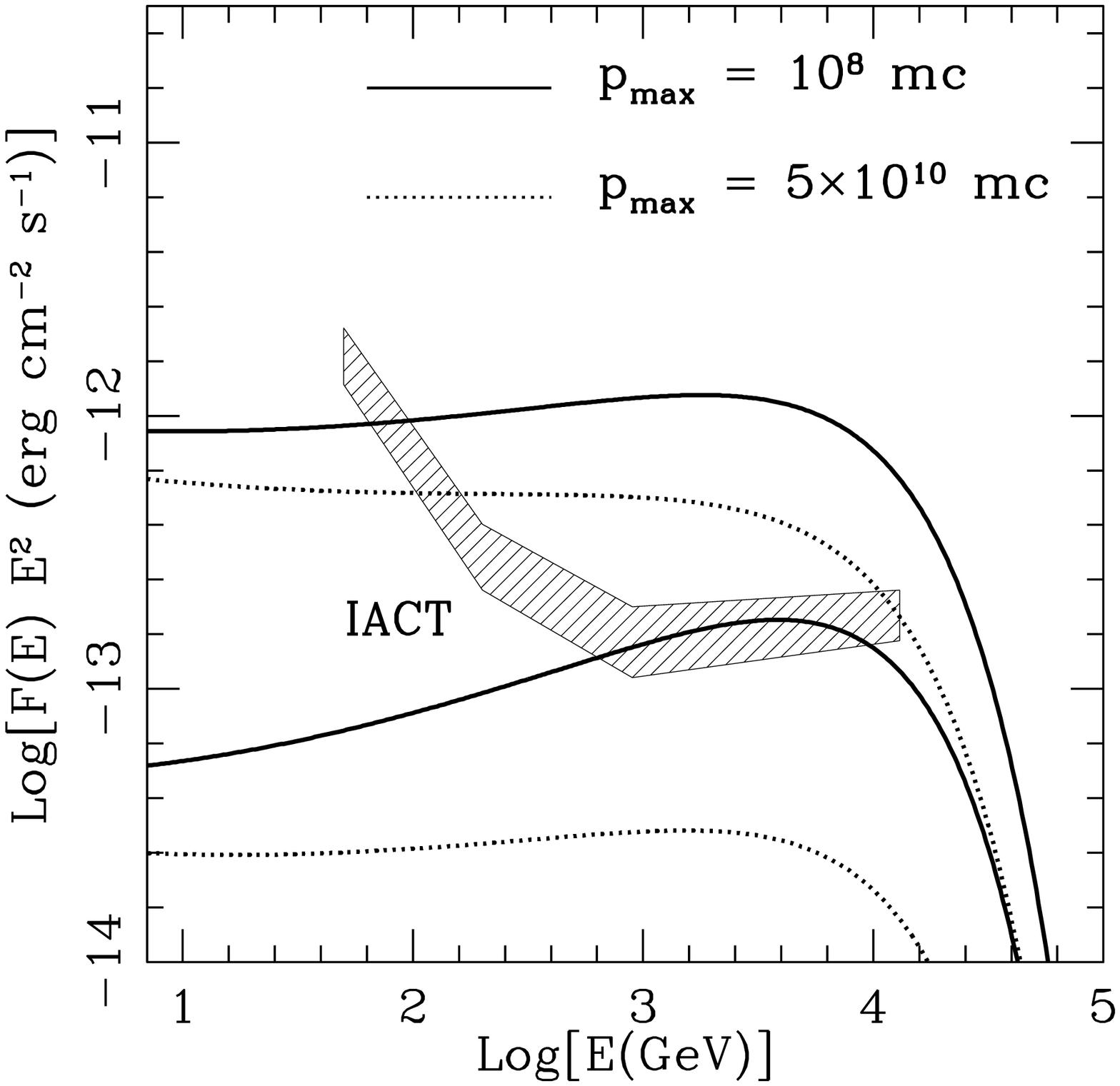}
\end{center}
\end{minipage}
\hfill
\caption{{\bf Left panel:} Spectra of protons accelerated at the accretion shock of a Coma--like cluster. The horizontal line represents the result of the linear theory (see text). {\bf Right panel:} Gamma ray fluxes for a Coma--like cluster compared with the HESS sensitivity for point sources. The external temperature is assumed to be $10^6 K$ (two upper curves) or $10^4 K$ (two lower curves).}
\end{figure}

Let us consider now a Coma--like cluster with mass $10^{15} M_{\odot}$ and at a distance of about $100 \, Mpc$. The spectra of particles accelerated at the accretion shock are shown in figure 2 (left panel). We assume here that the accretion shock is located at the cluster virial radius and that the infalling matter is moving at the free fall velocity $u_0 \sim 1700 \, km/s$. We consider different values for the Mach number (${\cal M} = 10$ and $100$) and for the maximum momentum of accelerated particles ($p_{max}/(mc) = 10^7$ and $5\times10^{10}$). As a comparison we plot the prediction of test--particle theory assuming that a fraction of $10\%$ of the shock kinetic energy flux is converted into CRs (orizontal lines). The differences between the two results are straightforward: non--linear spectra, when multiplied by $p^4$, exhibit a large dip with minimum at energyes of the order $\sim 100 - 1000 \, GeV$. This is exactly the energy range relevant for gamma ray production via proton--proton interactions. For this reason, the expected gamma ray emission might be suppressed with respect to earlier predictions, based on test--particle theory. 

A rough, order of magnitude estimate of the total amount of protons accelerated at the accretion shock and stored into the cluster can be given by $\epsilon f(p) u_2 \tau S$, where $f(p)$ is the particle distribution function at the shock, $u_2$ is the downstream velocity, $\tau$ is the cluster age, $S$ the shock surface and $\epsilon$ a factor that takes into account the effects of adiabatic compression. These protons can interact with the protons in the ICM, with typical density $n_{gas} \sim 10^{-4} cm^{-3}$ and produce gamma rays. 
Gamma ray spectra are plotted in Figure 2, together with the sensitivity of the HESS telescope for point sources. The effect of absorption of high energy photons in the cosmological infrared background has been taken into account.
For an extended sources of size $\sim 1^{\circ}$ the sensitivity curve should be roughly multiplied by a factor of $\sim 10$.
Since clusters have approximatively this size, it is clear that only for the most optimistic choice of parameters, these objects could be marginally detectable.
Of course, other contributions can increase the expected gamma ray luminosity, so that more accurate calculations and a accurate determination of the instrument sensitivity for extended sources are required.

\section{Conclusions and future perspectives}

In this paper we have discussed the issue of particle acceleration at large scale shocks in the ICM. We have shown that, due to the intrinsic efficiency of shock acceleration, the backreaction of the accelerated particles onto the shock structure has to be considered in order to obtain accurate results.
This is true in particular for high Mach number shocks, which are likely to convert a high fraction of the total shock kinetic energy into CRs.
As a consequence, the heating of the gas can be strongly suppressed at strong accretion shocks around clusters of galaxies and filaments.
Moreover, the spectra of accelerated particles are no longer power laws, being steeper at low energy (with respect to the $p^{-4}$ behaviour predicted by the test--particle theory) and flatter at higher energies, with an asymptotic slope $\alpha = 3/2$.
This spectral modification can lead to a suppression of the gamma ray emission and, in the most optimistic situation, the expected flux level might be roughly comparable with the HESS sensitivity.
However, an accurate determination of the instrument sensitivity for extended sources is needed.

Numerical simulations or semi--analytical calculations are commonly used to describe both the process of large scale structure formation and the acceleration of particles at shocks.
Both these approach have advantages and disadvantages and therefore they should be used in a complementary way in order to achieve optimal results.

The most important advantages of semi--analytical calculations is that they are much less time consuming than simulations, still remaining often acceptably accurate.
On the other hand, numerical simulation are undoubtedly more accurate and become necessary in order to describe the most complex situations.
This could be the case of cluster formation and evolution if CR pressure cannot be neglected.

For these reasons, we suggest that an approach based on a combination of high resolution cosmological simulations and semi--analytical models for non--linear shock acceleration might be the most effective way to study the impact of intracluster CRs onto the thermal and non--thermal properties of large scale structures.





\bibliographystyle{aipproc}   


\IfFileExists{\jobname.bbl}{}
 {\typeout{}
  \typeout{******************************************}
  \typeout{** Please run "bibtex \jobname" to optain}
  \typeout{** the bibliography and then re-run LaTeX}
  \typeout{** twice to fix the references!}
  \typeout{******************************************}
  \typeout{}
 }

\end{document}